\documentclass[intlimits,twoside,a4paper]{article}

\usepackage{pbox}
\usepackage{caption}
\usepackage{color}

\usepackage{amsmath}
\usepackage{amssymb}
\usepackage{graphicx}
\graphicspath{{figures_art1}}
\usepackage[T2A]{fontenc}
\usepackage[cp1251]{inputenc}

\usepackage[eqsecnum]{cmpj3}

\issue{2017}{20}{3}{33701}
\doinumber{10.5488/CMP.20.33701}

\title[From the bulk electrolyte solution to the electrochemical interface]%
{From the bulk electrolyte solution to the electrochemical interface}
\author[H. Cachet]{H. Cachet\refaddr{label1,label2} }
\addresses{
\addr{label1} CNRS, UMR 8235, Laboratoire Interfaces et Syst\`emes Electrochimiques, 4
Place Jussieu, F-75005 Paris, France
\addr{label2} Sorbonne Universit\'es, UPMC, Universit\'e Paris 06, UMR 8235, LISE, 4,
Place Jussieu, F-75005 Paris, France
}

\date{Received April 28, 2017, in final form June 12, 2017}

\begin{document}

\maketitle

\begin{abstract}
This paper is aiming at presenting some relevant contributions of
Jean-Pierre Badiali during the first ten years of his growing
scientific activity. This presentation does not contain new materials
but is based on a number of selected papers published in the seventies,
a part of them written in French. The presentation is organized around
three points. The first point, corresponding to his PhD thesis, is
concerned with the study of ion-solvent and ion-ion interactions in a
solution using complex dielectric permittivity measurements in the
Hertzian and microwave frequency range. The second one is concerned
with an analysis of the ion pair absorption band observed in the far
infrared region in terms of an interionic potential energy. The third
one is concerned with the metal-solution interface and his significant
advances on (i) the Lippmann equation linking electrocapillary and
electrical measurements and (ii) the contribution of the metal to the
differential interfacial capacity.
\pacs 71-10.Ay, 73-20.At, 77-22, 78-30.cd, 82-45.Fk
\end{abstract}

\section{Introduction}
{This paper is written in memory of Jean-Pierre
Badiali, providing a survey of his contribution during the first ten
years of his scientific activity. As a matter of fact, this paper does
not contain new results. It is based on the bibliography of J.P.
Badiali as given in the references. Initially, he had a general
training in electrical engineering without a particular predisposition
in the field of electrochemistry. In 1967, he entered the C.N.R.S.
research group n{\textdegree}4 {\textquotedblleft}Physics of Liquids
and Electrochemistry{\textquotedblright}, directed by Prof.~I.~Epelboin.
More specifically, he joined the team led by Dr. J.C.~Lestrade, with the main objective to investigate the dielectric
properties of electrolyte solutions, considered as partners of the
electrochemical interfaces. In the sixties and seventies, a large
number of laboratories around the world were invested in applying
dielectric spectroscopy for studying molecular motions in pure liquids,
which means measuring the complex dielectric permittivity in a large
frequency domain including microwaves.
In the case of electrolyte solutions, the number of laboratories involved was
much less, notably in UK the pioneer group led by J.B. Hasted \cite{1} and in Germany
those of R. Pottel \cite{2} and J. Barthel \cite{3}. A reason for that was
the experimental difficulty arising from the electrical conductivity of
ionic solutions. Thus, measurements below c.a. 100~MHz are dominated by
the losses due to the ionic conductivity preventing from reaching the
so-called static dielectric constant of the medium, contrary to
the case of pure liquids.
Another limiting factor at this time was the necessity to develop several
experimental set-ups (often one per frequency) to cover all the required frequency range.
When preparing his PhD thesis \cite{4}, his first
task was to complete the existing experimental set-ups covering the
frequency range, 0.1 to 10~GHz, by building a waveguide interferometer
functioning at 35~GHz. As it will be shown later, this extended
frequency domain was very useful to analyze the relaxation processes
arising from molecular and ionic motions, at the source of his first
theoretical derivations. The present paper is divided into three parts,
corresponding to three relevant contributions of J.P. Badiali. The
first one will be devoted to the modeling of ionic relaxation
process, as evidenced from spectroscopic measurements in organic
solvents with a low dielectric constant \cite{5,6,7,8,9,10}.
The second one concerns the emerging THz spectroscopy applied to highly
absorbing liquids \cite{11,12}, with
the observation of an absorption band in the far infrared region (20 to 150~cm$^{-1}$)
related to the vibration of a cation-anion pair and the interpretation
of the frequency corresponding to the absorption maximum in terms of an
interionic potential \cite{13,14,15}. The
third one will refer to the electrochemical interface, from the point
of view of surface tension and differential capacitance \cite{16,17,18,19,20}.
This was in the context of developing microscopic physical models of the
metal-ionic solution interface using a statistical mechanical treatment of the
whole interface \cite{21,22}.

\section{Dielectric spectroscopy of electrolyte solutions}

\subsection{Generalities}
As mentioned in the introduction, dielectric
spectroscopy is based on the measurements of the complex dielectric
permittivity
$\varepsilon^*=\varepsilon'-\ri \varepsilon''$ with ($\ri^2=-1$)
in a frequency range sufficiently wide to get a reliable
information both on the static properties of the medium (the so-called
dielectric constant) and on different relaxation processes arising
from molecular and ionic motions. It is worthwhile recalling that the
dielectric permittivity expresses the response in terms of current
density of a medium submitted to a perturbation of electric field. In
the case of pure polar liquids, in the Hertzian frequency range (a few
dozens of MHz to a few dozens of GHz), the observed response is that of
rotation of molecular species. When considering electrolyte solutions,
motion of charge carriers, as translational displacements, should now
be taken into account. On the basis of statistical mechanics
considerations, using the formalism of microscopic Maxwell equations, a
general expression for the complex dielectric permittivity
$\varepsilon^*(\omega)$ of an electrolyte solution was established as:
\begin{align}
  \varepsilon^*(\omega) = \varepsilon^*_{\text d}(\omega)+\frac{\sigma_0}{\ri\omega\varepsilon_0}
+ H^*(\omega),
\label{eq:epsstardew}
\end{align}
where $\sigma_0$ is the low frequency conductivity, $\omega$ is the angular frequency and
$\varepsilon_0=8.84\cdot 10^{-12}$~F/m \cite{7,8,9}.
Conductivity losses, expressed as $\sigma_0/(\ri\omega\varepsilon_0)$,
become a dominating contribution to $\varepsilon$
at low frequencies. The first term $\varepsilon_{\text d}^*(\omega)$
represents the response of molecular species. The third term
$H^*(\omega)$ represents the frequency dependent contribution of ions, including
cross effects between ions and molecular dipoles.

\subsection{Solution spectrum analogous to solvent spectrum}
This case corresponds to a charge transport
independent of frequency, which means only represented by a static
conductivity $\sigma_0$, the one which can be measured at audio frequencies. In this way, the
overall complex dielectric permittivity reads:
$\varepsilon^*= \varepsilon_{\text d}'+\ri [\varepsilon''_{\text d}+\sigma_0/(\ri\omega\varepsilon_0)]$,
where $\varepsilon_{\text d}'$ and $\varepsilon''_{\text d}$ only refer to the solvent
relaxation process. Such a situation is usually encountered for solutions
with solvents having a high dielectric
constant, for instance water, methanol, ethanol, N,N-dimethylformamide
(DMF) \cite{5,6}. Any change in the solvent relaxation parameters can be
attributed to the short range ion-solvent interactions. In a general
way, it has been established that the observed decrease of the solvent
static permittivity $\varepsilon_{\text s}(c)$, when increasing the salt concentration $c$,
provided a quantitative information on the ion solvation process. In
agreement with experiments, at low salt concentrations,
$\varepsilon_{\text s}(c)$ is a linear relation, with a slope~$\delta$.
The static solvent permittivity $\varepsilon_{\text s}$ is conveniently described by the Kirkwood-Fr\"ohlich expression:
\begin{align}
\varepsilon_{\text s}-\varepsilon_{\infty}=\frac{3\varepsilon_{\text s}}{2\varepsilon_{\text s}+\varepsilon_{\infty}}
\frac{N}{3kT\varepsilon_{0}}\left(\frac{\varepsilon_{\infty}+2}{3}\right)^{2}{g\mu^{2}},
\end{align}
in which $\varepsilon_\infty$
represents the electronic and atomic contribution to the permittivity,
$\mu$ is the dipole moment of a solvent molecule in vacuum, $k$ is the
Boltzmann constant, $T$ is the absolute temperature, $N$ is the number of solvent
molecules per volume unit participating in the polarization of
orientation and $g$ is the correlation factor between neighboring
solvent dipoles. Using the above formula, a linear relation for
$\varepsilon_{\text s}(c)$ yields a linear relation for $N(c)$ as $N(c) = N_0 -qc$
with the following expression for $q$:
\begin{align}
q=\frac{N_{{0}}\delta}{\varepsilon_{\text{s}0}}\frac{2\varepsilon_{\text{s}0}^{2}+\varepsilon_{\infty}^{2}}
  {\left(\varepsilon_{\text{s}0}-\varepsilon_{\infty}\right)\left(2\varepsilon_{\text{s}0}+\varepsilon_{\infty}\right)}\,.
\end{align}
As a result, $q$ is an estimate of a solvation
number defined as the number of rotation-blocked solvent molecules per
molecule of salt. It has been shown that for cations such as
Li$^+$, Na$^+$, Mg$^{2+}$,
the short range cation solvent molecule interactions could be
coordinated bonds formation at the number of 4 per cation, in a
tetrahedral configuration. For symmetry reason, the solvated cations
bear no permanent dipole moment. Anions such as ClO$_4^-$
are assumed to be not solvated and small anions such as halides may interact
with molecules but without building well defined non-polar structures.
A number of significant data are given in table~\ref{tab1}, showing that the
number of blocked molecules are found to be very high in hydrogen bonded
solvents due to their self-association. In alcohols, the solvation
shell is not limited to the molecules coordinated with the cation but
involves rigid molecular chains, the mean number of molecules per chain
being given by the quantity $f$. For a simple liquid such as DMF without
dipolar correlations, $f$ is close to 1, the number of blocked molecules
is close to 4, as predicted by the cation solvation model.
\begin{table}[!t]
\renewcommand{\arraystretch}{1.2}
\caption{Effect of solvent intermolecular structure on the short range ion-solvent interactions
as measured by the solvation number $q$.} \label{tab1}
\vspace{2ex}
\begin{center}{\footnotesize  
\begin{tabular}{|c|c|c|c|c|c|c|c|}
\hline\hline
  & \multicolumn{6}{c|}{Self-associated protic solvents} & \pbox{3cm}{Non-associated\\ aprotic solvent} \\
  \hline\hline
solvent & \multicolumn{3}{c|}{Ethanol} & \multicolumn{3}{c|}{Methanol} & DMF \\
  \hline
Salt & LiClO$_4$ & Mg(ClO$_4$)$_2$ & LiCl & LiCLO$_4$ & NaClO$_4$ & Mg(ClO$_4$)$_2$ & NaClO$_4$ \\
  \hline
$\delta$, mol$^{-1}\cdot$ L & 25.4 $\pm$ 9.6 & 20.7 $\pm$ 6.6 & 26.6 $\pm$ 2.5 & 38.6 $\pm$ 3.9 & 31.5 $\pm$ 5.3& 32.7 $\pm$ 5.9 & 14.4 $\pm$ 2.6 \\
  \hline
$q$ & 18.5 $\pm$ 7 & 15.1 $\pm$ 4.8 & 19.5 $\pm$ 1.8 & 30 $\pm$ 3.1 & 27 $\pm$ 4.2 & 25.8 $\pm$ 4.7 & 4.6 $\pm$ 0.9 \\
  \hline
 $f$ & \multicolumn{3}{c|}{5 $\pm$ 1} & \multicolumn{3}{c|}{7 $\pm$ 1} & $\approx 1$ \\
  \hline\hline
\end{tabular}}
\end{center}
\end{table}

\subsection{Solution spectrum with additional ionic relaxation process}
Dielectric experiments of ionic solutions in weakly polar solvents, such as
acetone, ethyl acetate, tetrahydrofuran (THF), clearly evidence a new
relaxation process in addition to that of the solvent \cite{7,10}. A first
attempt in literature to interpret this phenomenon was to consider the ion
pair formation, in a way analogous  to what was done in classical
theories of conductivity of dilute electrolytes. Ion pairing is based
on the static concept of a chemical equilibrium between ion pairs and
free ionic species. Neutral ion pairs do not
participate in charge transport and are capable of rotating as molecular
dipoles with a definite dipole moment. Such an approach gives rise to
several objections: (i) at relatively high salt concentrations,
necessary to extract the ionic contribution from dielectric
experiments, ionic aggregates of the orders  higher than ion pairs should be
introduced, as done in conductivity theories; (ii) ion pair cannot be
actually considered as having an infinite life-time. To clarify these
objections, Badiali et al. derive a formal theoretical expression
for the conductivity and its frequency dependence which will be briefly
summarized below \cite{7}. It has been shown that the time dependence of the
current density $I(t)$ could be written as:
\begin{align}
   I(t)=\frac{1}{3kTV}
    \int_{0}^{t} \langle \dot{M}(0)\dot{M}(\tau)\rangle E(t-\tau ) \rd\tau.
\label{eq:Idet}
\end{align}
In equation~(\ref{eq:Idet}), $\dot{M}$ is the time
derivative of the electrical moment $M$ of all
the entities inside the volume~$V$, the neutral ones (solvent molecules)
with moment $M_{\text m}$, the charged species (ions, solvated or not, other charged aggregates)
with moment $M_{\text c}$, with $M=M_{\text m}+P_{\text c}$.
This decomposition of $M$ when inserted into equation~(\ref{eq:Idet}) gives rise to three time
dependent current contributions, (i) molecular relaxation; (ii) ionic
relaxation leading to a frequency dependent conductivity $\sigma(\omega)$;
(iii) coupling between dipolar and charged species usually considered
as negligible. The following expression was derived for the frequency
dependent conductivity $\sigma(\omega)$:
\begin{align}
\sigma (\omega )=\sigma (0)+\ri\omega\frac{1}{3kTV}
   \langle M_{\text c}^{2}(0)\rangle \left[1-\ri\omega \int^{\infty}_{0}
  \re^{-\ri\omega \tau}\frac{\langle M_{\text c}(0)M_{\text c}(\tau)\rangle}{M_{\text c}^{2}(0)} \mathit{\rd\tau } \right].
\end{align}

Then, the spectral response of electrolytes is
connected with the self-correlation function of the electrical moment
of all the ions in a given volume. Using the above expression, a
statistical model of charge transport was derived considering around an
anion the relative motion of a cation perturbed by collisions with
other species with a probability $p$ to be weak, $(1-p)$ to be strong, and
a linear motion between collisions. The collision times are assumed to
be distributed according to a Poisson process. It leads to a class of
non-exponential correlation functions as suggested by experiments. In
the special case of a linear Brownian motion between collisions, an
analytical expression was obtained for the ionic conductivity $\sigma(\omega)$.
It can be expressed in terms of a complex permittivity
$\varepsilon_{\text c}(\omega)$:
\begin{align}
\varepsilon_{\text c}(\omega)=\frac{\sigma(\omega)}{\ri\omega \varepsilon_{0}}=
\frac{\sigma(0)}{\ri\omega \varepsilon_{0}}+
\frac{\Delta \varepsilon}{1+\ri\omega \tau}
\left(1+\gamma \frac{\ri\omega \tau}{\sqrt{1+a+\ri\omega \tau}}\right).
\label{eq:epscdew}
\end{align}

The second term in equation~(\ref{eq:epscdew}) provides an analytical
expression to the $H^*(\omega)$
function appearing in the general expression of the response
$\varepsilon^*(\omega)$
of an electrolyte solution (\ref{eq:epsstardew}). This model was successful for
representing the experimental data of electrolyte solutions in weakly
polar solvents. As an example, figure~\ref{fig:dielfreq} shows the variations of the real
and imaginary parts of $\varepsilon^*(\omega)$
with frequency in the case of solutions of LiClO$_4$
in tetrahydrofuran at 30{\textdegree}C. The agreement between model and
experiments is of the order of 1\% on the real and imaginary parts of
$\varepsilon^*(\omega)$.

\begin{figure}[!b]
\begin{center}
\includegraphics[width=0.46\textwidth]{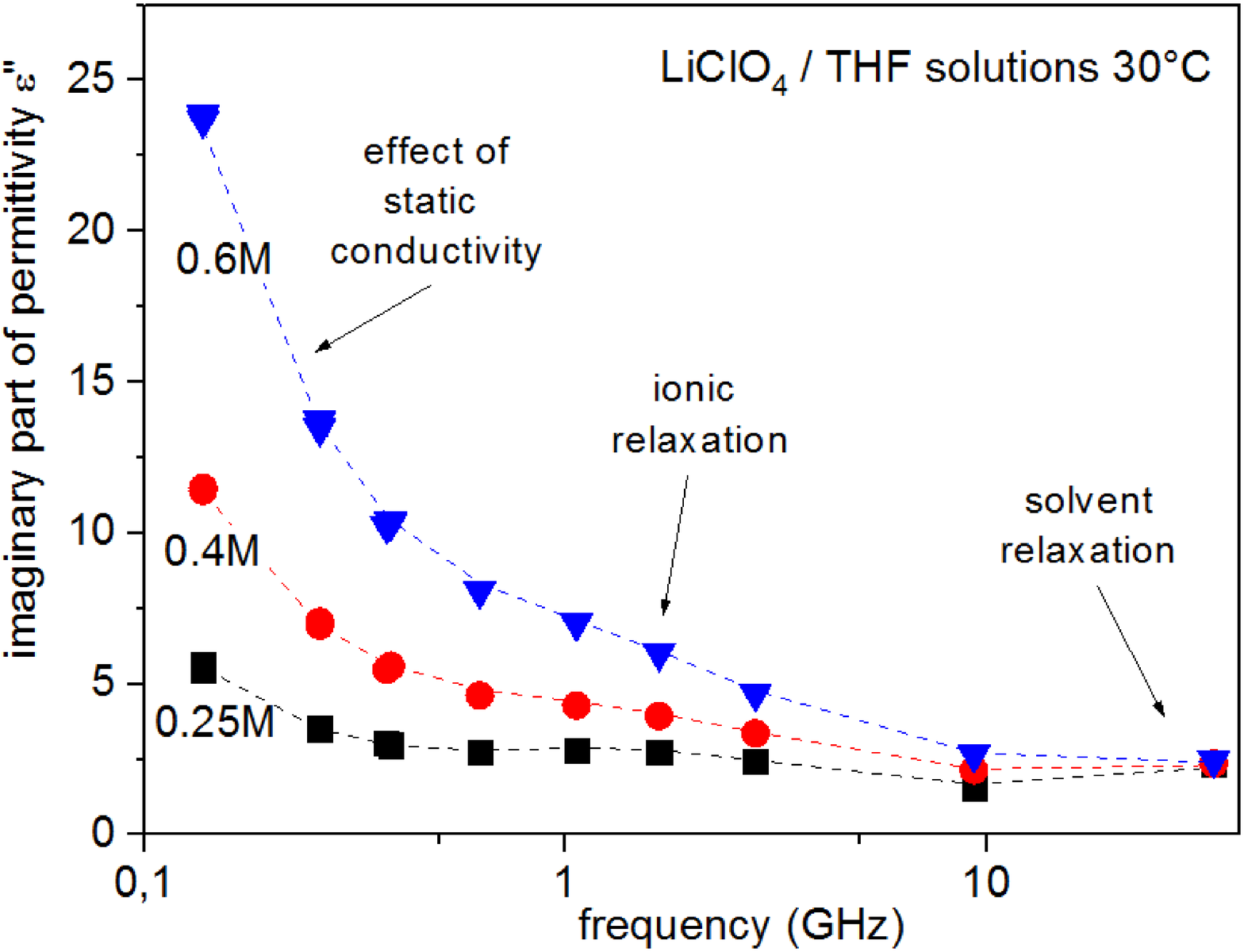}\hspace{2ex}
\includegraphics[width=0.46\textwidth]{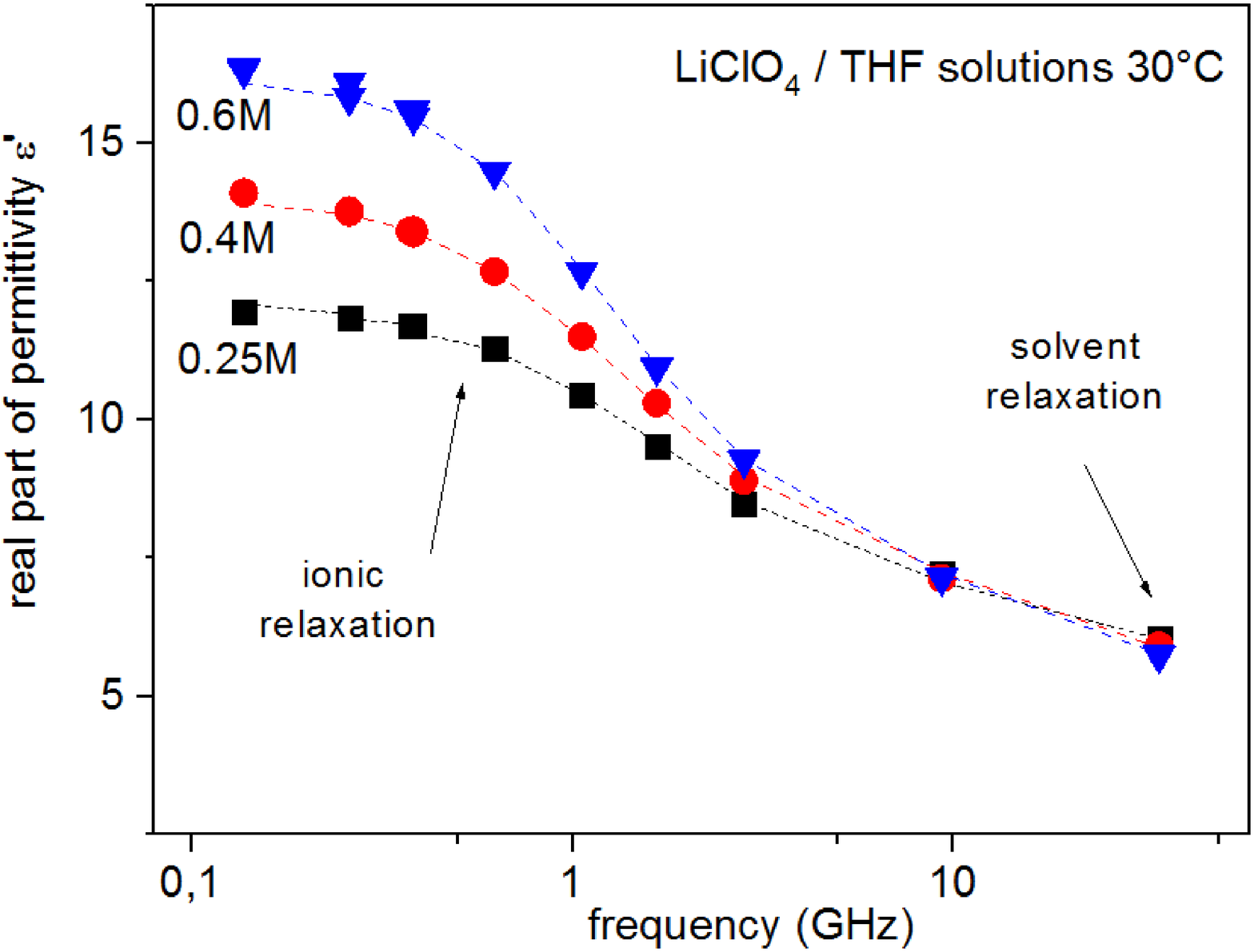}
\caption{(Color online) Variations with frequency of the real and imaginary parts of $\varepsilon^*(\omega)$ for
LiClO$_4$/tetrahydrofuran solutions at 30$^{\circ}$C: ({\small$\blacksquare$}) 0.25M; (\textcolor{red}{\Large$\bullet$}) 0.4M; (\textcolor{blue}{$\blacktriangledown$}) 0.6M. The lines
correspond to the numerical fit of the frequency dependent conductivity model completed by
the contribution of solvent relaxation of the Debye type (relaxation time: $\tau_{\text d}$ = 2.3~ps;
dielectric constant: 7.22 at 30$^{\circ}$C).}
\label{fig:dielfreq}
\end{center}
\end{figure}

\begin{table}[!t]
\begin{center}
\renewcommand{\arraystretch}{1.2}
\caption{Parameters extracted from the numerical analysis of the
permittivity data of LiClO$_4$/THF solutions according to the full expression
given at the top of the table; $\varepsilon_\infty$ ($= 2.19$) and $\tau_{\text d}$ ($=2.3$~ps)
were fixed at their values for the pure solvent.} \label{tab2}
\vspace{2ex}
\begin{tabular}{|c|c|c|c|c|c|c|c|}
\hline\hline
  \multicolumn{8}{|c|}{$\displaystyle \varepsilon_{\text c}(\omega)=
\frac{\sigma(0)}{\ri\omega \varepsilon_{0}}+
\frac{\Delta \varepsilon}{1+\ri\omega \tau}
\left(1+\gamma \frac{\ri\varpi \tau}{\sqrt{1+a+\ri\omega \tau}}\right)+
\frac{\varepsilon_{\text s}-\varepsilon_\infty}{1+\ri\omega\tau_{\text d}}+\varepsilon_\infty$}\\
  \hline\vspace{-2.5mm}
Salt molar  & \raisebox{-0.2cm}{$\varepsilon_{\text s}$} & \raisebox{-2mm}{$\Delta\varepsilon$} & \raisebox{-2mm}{$\sqrt{\langle R^2(0)\rangle}$, nm} &
   \raisebox{-2mm}{$\tau$, ps} & \raisebox{-2mm}{$p=a/(1+a)$} & \raisebox{-2mm}{$\sigma_0$, S/m} & \raisebox{-2mm}{Fit quality}\\ 
 concentration &&&&&&&\\  
  \hline \hline
0.25 & 6.72 & 5.3 & 0.38 & 126 & 0.32 & 0.039 & 1.04\\
  \hline
0.4 & 6.14 & 7.9 & 0.37 & 194 & 0.43 & 0.081 & 0.91\\
  \hline
0.6 & 5.97 & 10.2 & 0.34 & 136 & 0.65 & 0.171 & 1.04\\
  \hline\hline
\end{tabular}
\end{center}
\end{table}

Table~\ref{tab2} shows the full expression of $\varepsilon^*(\omega)$
used to represent the experimental spectra and the values of the
relevant parameters obtained by numerical fitting. As previously, the
static dielectric constant of the solvent $\varepsilon_{\text s}$
decreases at an increasing salt concentration, in agreement with a
solvation number close to~4. The characteristic time of relaxation of
ionic atmosphere $\tau$
lies in the range 130--200~ps,
one hundred times larger than the dipole rotation time. The probability
factor $p$ for the occurrence of weak collisions turns around 0.5. The
mean anion-cation quadratic distance
	$\delta_{\text{ac}} = \sqrt{\langle R^2(0)\rangle }$
can be calculated from the amplitude $\Delta \varepsilon$
of the ionic relaxation according to:
\begin{align}
\sqrt{\langle R^{{2}}(0)\rangle }=\frac{3kT\varepsilon _{0}}{e^{2}}
\frac{\Delta \varepsilon}{N/V}\,,
\end{align}
where $e$ is the electronic charge. The values of
$\delta_{\text{ac}}$
are slightly higher than the sum of the crystallographic radii (0.26~nm),
in good agreement with the fact that the cation is solvated and
that the ion pair is a fluctuating labile structure.

\section{Interionic potential energy and far-infrared signature}
As mentioned above, the microwave properties
of electrolyte solutions in non-aqueous solvents consist mainly in an
excess dielectric polarization and absorption which is interpreted in
terms of ionic processes at the time scale of about $10^{-9}$~s.
A realistic interpretation of this phenomenon was proposed, based on a
stochastic description of the translational diffusion of ions,
perturbed by instantaneous collisions. In the far-infrared range (FIR),
at millimeter and submillimeter wavelengths (10 to 150~cm$^{-1}$
wave numbers), molecular liquids present an absorption originating from
rotational motion, combined eventually with low frequency
vibration modes. In the presence of ions,
additional effects can occur due to ion-ion and ion-solvent
interactions, each one depending on the nature of the solvent and ions.
In what follows, we are dealing with the FIR response of
tetraalkylammonium halides R$_4$NX
for which a model for the ion-ion pair potential energy was elaborated.
FIR spectra of R$_4$NX
in solvents such as benzene, carbon tetrachloride and chloroform are
characterized by a very broad and asymmetric absorption band, the
frequency position $\nu_{\text L}$ being anion dependent, as depicted in figure~\ref{fig:FIR} \cite{13,14,15}.

This is in agreement with the harmonic oscillator model for the anion-cation pair,
the resonant frequency $\nu_{\text L}$
being related to the force constant $k_{\text L}$:
\begin{align}
\nu_{\text {L}}=\frac{1}{2\piup c_0}\sqrt{\frac{k_{\text L}}{m_{+-}}}\,,
\end{align}
where $c_0$ is the light velocity and $m_{+-}$
is the reduced mass of the ion pair, close to the anion
mass because the cation is much heavier than the anion.
It was shown that a realistic
effective pair potential $u_{ij}$
could be derived from properties of the salt in the solid state. The
proposed form was:

\begin{align}
u_{ij}(r_{ij})=\frac{1}{4\piup \varepsilon_{0}}
 \left[\frac{e_{i}e_{j}}{r_{ij}}-
\frac{(\alpha_{i}+\alpha_{j})e^{2}}{2r_{ij}^{4}}-
\frac{c_{ij} e^{2}}{r_{ij}^{6}}+
\frac{be^{2}}{r_{ij}^{9}}\right],
\end{align}
where $r_{ij}$
is the distance between the ions $i$ and $j$, having a charge $e_i$ or $e_j$,
and polarizabilities $\alpha_i$ or $\alpha_j$, respectively. If $r_{\text L}$
is the equilibrium distance between the centers of both charges, the
force constant $k_{\text L}$ governing the band position is given by:
\begin{align}
k_{\text L}=\frac{e^{2}}{4\piup \varepsilon_{0}}
\left(\frac{90b}{r_\text{L}^{11}}-\frac{2}{r_\text{L}^{3}}-
\frac{20\alpha}{r_\text{L}^{6}}-\frac{42c}{r_\text{L}^{8}}\right).
\end{align}

\begin{figure}[!t]
\begin{center}
\vspace{-4mm}
\includegraphics[width=0.4\textwidth]{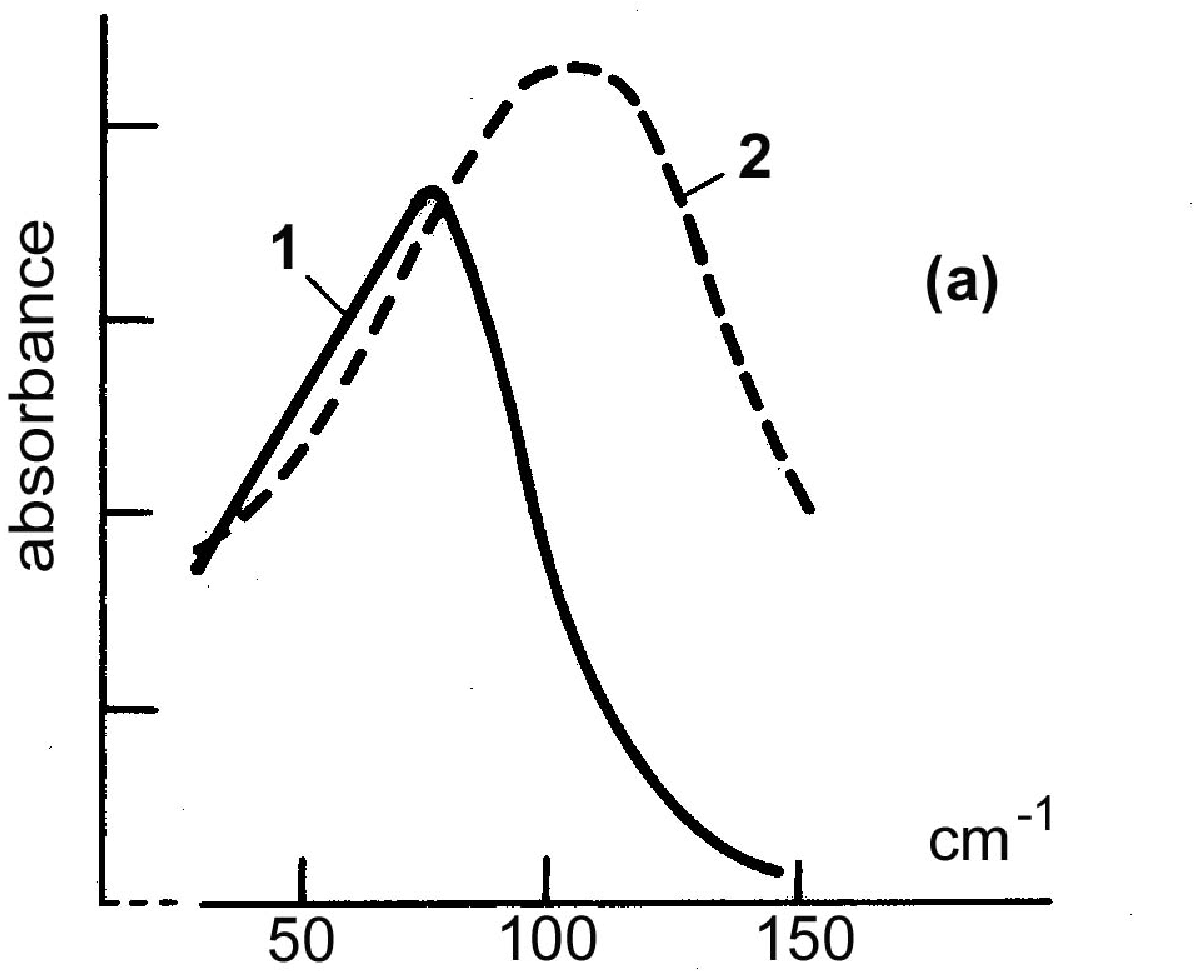}\hspace{2ex}
\includegraphics[width=0.48\textwidth]{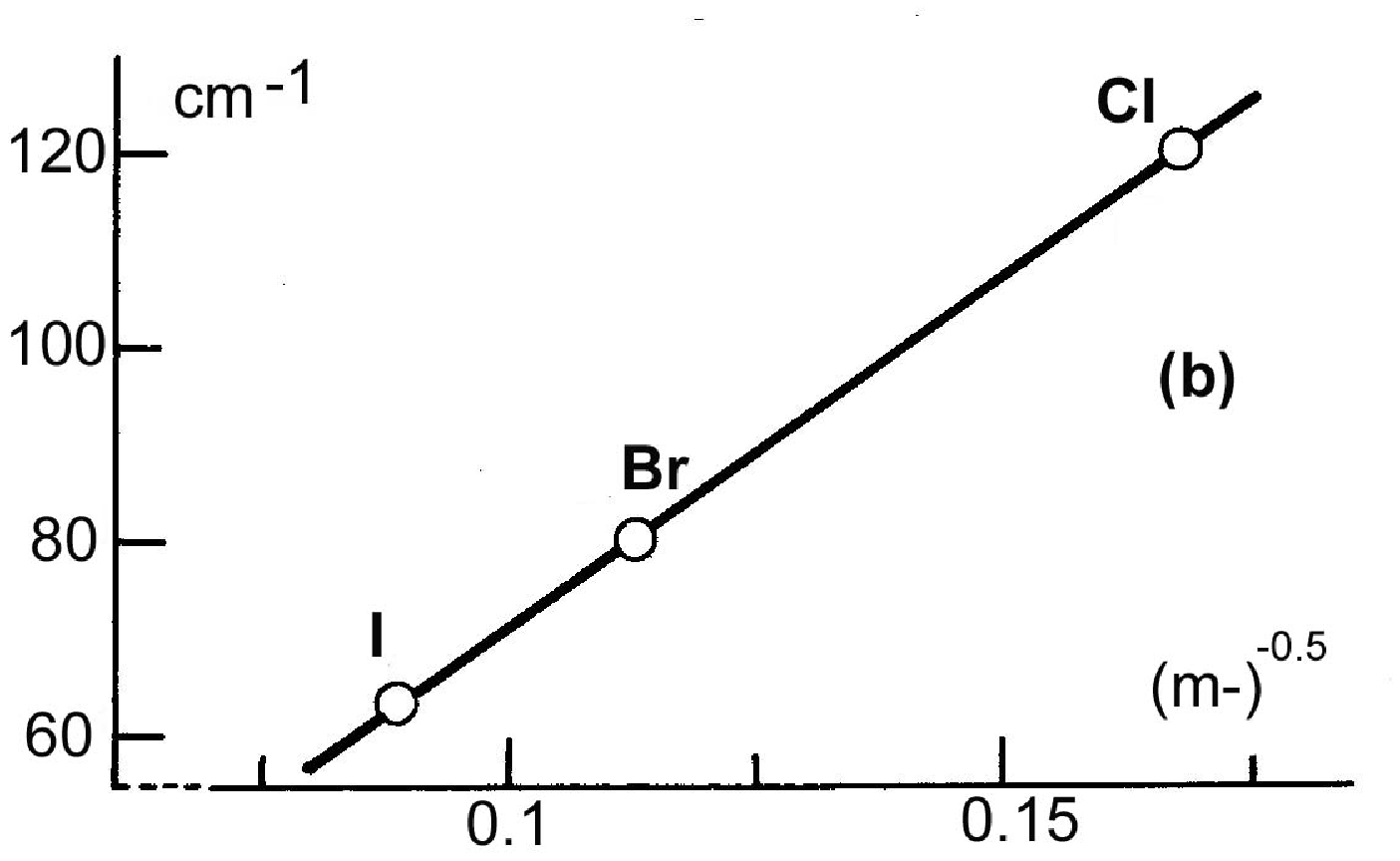}
\caption{FIR behavior of tetra-alkylammonium halides: (a) absorption spectra of Pr$_4$NBr in
chloroform~(1); Bu$_4$NCl in a 5\%~v/v acetone/CCl$_4$; (b) dependence of the maximum
absorption frequency on the anion mass $m_-$.}
\label{fig:FIR}
\end{center}
\end{figure}

In this expression, $b$ is determined from the properties of the solid.
Assuming a body-centered cubic arrangement (CsCl type) for
the structure of R$_4$NX
when unknown leads to calculated frequencies in satisfying agreement
for the solid salts but lower than the experimental ones in solution. A
more refined calculation was done in the case of
tetrapropylammonium bromide Pr$_4$NBr
for which the actual structure is known as a zinc sulfide type
arrangement. Due to a tetragonal structure, there is a distribution
of frequencies according to the crystallographic directions. As shown
in table~\ref{tab3}, the $b$ parameter relative to the repulsive term is strongly
dependent on the structure, and the right structure leads to a fine
agreement between experimental and calculated $\nu_{\text L}$
frequencies, validating the physical image of an ion pair vibrator in
solution at the $10^{-12}$--$10^{-13}$~s time scale.

\begin{table}[!t]
\begin{center}
\renewcommand{\arraystretch}{1.2}
\caption{Calculated and experimental frequencies $\nu_{\text S}$ (solid) and $\nu_{\text L}$ (in solution) for
Pr$_4$NBr.} \label{tab3}
\vspace{2ex}
\begin{tabular}{|c|c|c|c|}
\hline\hline
  \multicolumn{4}{|c|}{Pr$_4$NBr} \\
  \hline\hline
Structure & $10^{76}\cdot b$, m$^8$ & $\nu_{\text S}$, cm$^{-1}$ & $\nu_{\text L}$, cm$^{-1}$ \\
  \hline
Cubic & 25.5 & 56 & 55\\
  \hline
Tetragonal &  7.2 & 85 to 98 & 78\\
  \hline
Experimental values & & 48-71-104 & 80\\
  \hline\hline
\end{tabular}
\end{center}
\end{table}

\section{Electrochemical interface}
 Among the possible electrochemical systems, the ideally polarisable
electrochemical interface is the simplest one, in the sense that no
charge transfer occurs across it. The most widely studied example is
given by the mercury drop electrode, which allows one to perform
electrocapillary and electrical impedance measurements. Such an ideal
system was an attractive field for Badiali to develop a general
statistical mechanical treatment in comparison with the usual
thermodynamic approach. The first important contribution was concerned
with the notion of surface tension, more precisely with the Lippmann
equation, the second one was concerned with the contribution of the
metal to the differential capacitance of the ideally polarized
electrode.

\subsection{The Lippmann equation}
The Lippmann equation is concerned with
the surface tension $\gamma$ of the
interface between an ideally polarisable electrode and an ionic
solution. According to this equation, the change in surface tension
$\Delta \gamma$,
divided by the change in the potential drop across the interface
$\Delta U$,
gives the negative of the surface charge density $Q$. As a consequence,
the second derivative of the surface tension with potential can be
identified with the capacity of the double layer $C_{\text{dl}}$
obtained from impedance measurements. It reads:
\begin{align}
\Delta \gamma /\Delta U=-Q \qquad  \mbox{and}\qquad
-\partial^{2}\gamma /\partial^{2}U=C_{\text{dl}}\,.
\end{align}
When derived by thermodynamics, the quantities
appearing in the Lippmann equation do not refer to the actual charge
distribution in the interfacial region. The original work of Badiali
and Goodisman was to derive the Lippmann equation by statistical
mechanical methods on the basis of a model at the molecular level,
enabling all the quantities appearing in the model to have a physical
definition \cite{16,17}. The first step was to derive the conditions for
mechanical equilibrium in the presence of an electric field of a system
with inhomogeneous and anisotropic properties. From the balance of
forces, equations were obtained for the surface tension in terms of the
pressure, electric field, electric charge density, and electric
polarization at each point within the system. Having in mind the
mercury drop electrode, a spherically symmetric system was considered,
allowing a direct calculation of the change in the surface tension
$\Delta \gamma$ produced by a change in the potential drop $\Delta U$,
maintaining thermal equilibrium, constant temperature, and the pressure
and chemical composition in homogeneous regions.
A surface on which the charge density is always zero was introduced
within the interface,
allowing the surface charge to be defined as the integral of the charge
density over the metal side of the interface. Only the solution side
was treated by statistical mechanics using Boltzmann distributions for
charged and polarisable species. The Lippmann equation was successfully
derived in two cases: (i) considering only ions and assuming a
dielectric constant equal to that of vacuum; (ii) considering ions and
molecules in thermal equilibrium, and a dielectric constant varying
from a point to point and changing with the field. It was emphasized that, in
a coherent model, it was inadmissible of reducing the solvent to a
medium of fixed dielectric constant. Finally, considering the response
of the system to an imposed alternating potential, within the low
frequency limit, it was demonstrated that the system behaves as a pure
capacitance with a value equal to the derivative of the above mentioned
surface charge density with respect to the potential drop across the
interface.

\subsection{Contribution of the metal to the
interfacial differential capacity}
Experimentally, it was found that the
point of zero charge (pzc) and the interfacial differential capacity of
a metal-solution interface are dependent on the metal. To understand this
dependence, Badiali et al.
 firstly intended to calculate how the
surface potential can be modified by the metal solvent interaction at
the pzc \cite{18}, and subsequently with the charge on the metal in view of
calculating the interfacial capacitance \cite{19,20}. The model for the
metal surface in contact with a solvent they retained was the so-called ``dielectric field model''
represented in figure~\ref{fig:dielmodel} with parameters for an aqueous solution.
Assuming no specific adsorption, the metal should be in contact with a
monolayer of water, whose thickness
$d_2-d_1$
is about 0.3~nm and attributing a relative dielectric permittivity
$\varepsilon_1=6$ for the fixed dipoles. It is assumed that the water monolayer is
bounded by an ideal charged plane (located at $d_2$),
beyond which one has an unperturbed solvent with dielectric
permittivity $\varepsilon_1=78$.
The distance $d_1$ is the distance of the closest approach between the
metal ion cores and the center of adsorbed solvent molecules.
A calculation of $d_1$ was proposed by Kornyshev and Vorotyntsev, about 0.13~nm
in the case of mercury \cite{23}.
For Badiali et~al., $d_1$
was considered as an equilibrium distance as the result of
metal-solution coupling, which is charge dependent.

\begin{figure}[!t]
\begin{center}
\includegraphics[width=0.65\textwidth]{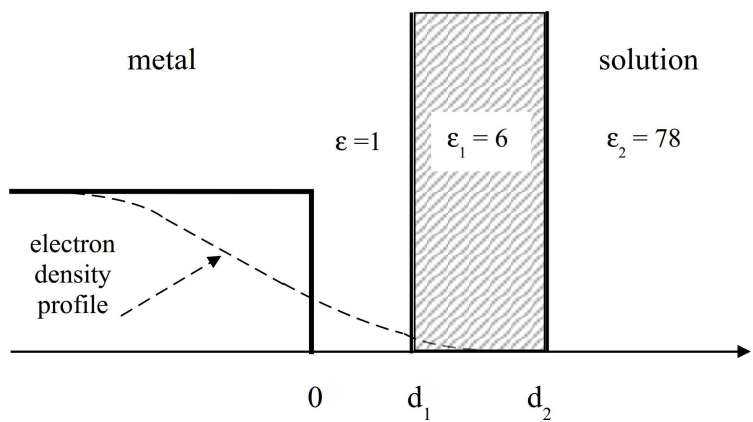}
\caption{Dielectric field model used to discuss the interaction of solvent molecules with metal
electrons.}
\label{fig:dielmodel}
\end{center}
\end{figure}

Using a two-parameter exponential profile for electrons, the above model
leads to an expression of the interfacial differential capacity $C$ at
the pzc:

\begin{align}
\frac{1}{C}=\frac{d_{2}-d_{1}}{\varepsilon_{0}\varepsilon_{1}}+
\frac{d_{1}}{\varepsilon_{0}}+\frac{1}{C^{\text M}(\text{dip})}\,.
\end{align}

$C^{\text M}(\text{dip})$ corresponds to the relaxation of the electronic structure with the
charge on the metal. This approach has successfully taken into account
the large difference in the capacitance values obtained for a mercury
electrode $C(\text{Hg}) = 28$~F/cm$^2$,
and for a gallium electrode $C(\text{Ga}) = 135$~F/cm$^2$,
emphasizing the crucial role played by the metal in determining the
value of the interfacial capacity.

\section{Conclusion and perspectives}
	The research works presented in this paper reflect a certain state of the art
forty or fifty years ago. A number of progresses have been made since that time.
This is particularly the case for the ultra-broadband dielectric spectroscopy
(1~MHz to 10~THz), mainly due to technological developments, although several
different apparatuses with limited bandwidth remain still required for covering
the full frequency range. For instance, by the use of network analyzers it
becomes possible to record a dielectric spectrum over three decades within a
minute instead of hours \cite{24}. The advantages and capabilities of modern
dielectric spectroscopy for investigating ion-ion and ion-solvent interactions
can be found in the review paper of Buchner and Hefter \cite{25}. For the
electrochemical interface, the 1980s were a very fruitful period through the
development of truly molecular models for the whole interphase based on jellium
for the metal and on ensembles of hard spheres for the electrolyte solution.
Such a mathematical description of the metal-solution interface, in which
Badiali was an active contributor, was progressively completed by numerical
experiments of increasing complexity as stated in \cite{26}.
As an example, such theoretical experiments were performed to describe the branching
pattern of the capacitance-voltage curves at the metal --- ionic solution interface \cite{27}.
The tendency in
recent works is the shift from aqueous systems to ionic liquids to understand
the bell or camel-shaped capacitance-potential curve around the point of zero
charge, not explainable within the classical Gouy, Chapman and Stern theory \cite{28}.

From the above review, it is clear that J.P. Badiali was early on
interested in providing a physical explanation of a number of
experimental facts in the field of electrochemistry, using fundamental
theoretical approaches. This attitude was always his during his brilliant
career of researcher.

\ukrainianpart

\title{Від об'ємного розчину електроліту до електрохімічного інтерфейсу}
\author{Г. Каше\refaddr{label1,label2} }
\addresses{
\addr{label1}  Національний центр наукових досліджень Франції, Лабораторія інтерфейсів  та електрохімічних систем, Париж, Франція
\addr{label2} Університет Сорбонна, Університет імені П'єра та Марії Кюрі, Університет м. Париж,   Париж, Франція
}

\makeukrtitle

\begin{abstract}

Метою статті є представлення важливого внеску Жана-П'єра Бадіалі впродовж перших десяти років його стрімкої наукової діяльності. Вона не містить
нового матеріалу, проте базується на низці вибраних статтей, опублікованих в 70-х роках, частина з яких написана французькою мовою. Стаття представляє три аспекти. Перший, що відповідає тематиці дисертації доктора філософії, присвячений  вивченню взаємодій ``іон-розчинник'' та ``іон-іон'' в розчині з використанням складних  вимірювань діелектричної проникності в  діапазоні  Герців і мікрохвильових  частот. Другий аспект стосується аналізу зони поглинання іонних пар, яка спостерігається в далекій інфрачервоній області в термінах міжіонної потенціальної енергії. Третій пов'язаний з поверхнею розділу ``метал-розчин'', і  важливими досягненнями тут є: (і) рівняння Ліппмана, яке пов'язує
електрокапілярні  та електричні вимірювання; (іі) внесок металу в диференціальну інтерфейсну електричну ємність.

\keywords діелектрична спектроскопія, іонна провідність, поверхневий натяг, інтерфейсна ємність
\end{abstract}

\end{document}